\documentclass[11pt,twoside]{article}
\usepackage{asp2010}

\resetcounters

\begin{document}

\title{Rule-based Cross-matching of Very Large Catalogs in NED}
\author{Patrick~M.~Ogle, Joe~Mazzarella, Rick~Ebert, Dario~Fadda, Tak~Lo, Scott~Terek, Marion~Schmitz, and the NED Team
\affil{Infrared Processing and Analysis Center, California Institute of Technology, 1200 East California Boulevard, Pasadena, CA 91125, USA}
}

\begin{abstract}

The NASA/IPAC Extragalactic Database (NED) has deployed a new rule-based cross-matching algorithm called Match Expert (MatchEx), capable of cross-matching
very large catalogs (VLCs) with $>10$ million objects. MatchEx goes beyond traditional position-based cross-matching algorithms by using
other available data together with expert logic to determine which candidate match is the best.  Furthermore, the local background
density of sources is used to determine and minimize the false-positive match rate and to estimate match completeness. The
logical outcome and statistical probability of each match decision is stored in the database, and may be used to tune the algorithm and
adjust match parameter thresholds. For our first production run, we cross-matched the GALEX All Sky Survey Catalog (GASC), containing
nearly 40 million NUV-detected sources, against a directory of 180 million objects in NED. Candidate matches were identified for each
GASC source within a $7\farcs5$ radius. These candidates were filtered on position-based matching probability, and on other criteria
including object type and object name. We estimate a match completeness of 97.6\% and a match accuracy of 99.75\%. MatchEx
is being used to cross-match over 2 billion catalog sources to NED, including the Spitzer Source List, the 2MASS Point-Source Catalog,
AllWISE, and SDSS DR 10. It will also speed up routine cross-matching of sources as part of the NED literature pipeline.

\end{abstract}

\section{Introduction} 

\subsection{Background}

The NASA/IPAC Extragalactic Database (NED) provides a comprehensive fusion of multiwavelength data on extragalactic
astrophysical objects, from data published in the astronomical literature and large, online catalogs \citep{hm88,h90,m07}. 
With a rapid increase in data volume from space and ground-based surveys, NED is developing new methods for keeping apace.
We have implemented a new rule-based cross-matching algorithm for very large catalogs (VLCs) with $>1\times10^7$ sources.
All catalog source and NED object attributes, continuous (e.g. position) or discrete (e.g. object type), are 
potential match discriminants.  Match metrics and outcome are tabulated in the database for statistical analysis. 
We chose the 40 million source GALEX All Sky Survey Catalog as the first VLC for NED to tackle with this methodology. NED will
continue to ingest and integrate even larger VLCs over the next few years, including the the Spitzer Source List, the 2MASS Point
Source Catalog, AllWISE, and SDSS DR10.

Cross-matching is a central component of the Virtual Observatory concept because it is a prerequisite for combining
multiwavelength data \citep{mdk12}. One important application is the identification of rare objects of interest by their SEDs, for 
example brown dwarfs selected by their SDSS-2MASS colors \citep{mkb08, gsk11}. Methods for cross-matching VLCs typically involve selecting
candidate matches between two catalogs within a pre-defined separation threshold.  Other algorithms utilize Bayesian statistics to select the
most likely match, based on any number of parameters \citep{bs08}. We have opted for an approach that begins with positional matching then
applies additional criteria to select among match candidates, making use of the rich array of parameters available in NED. We measure the local
density of background objects in the vicinity of each source to estimate the Poisson likelihood that an object is either a good match or a
background object.

For NED, we make a distinction between entries in an incoming catalog (catalog sources) and 
the distinct entries in the NED object directory that we match them to (NED objects). Catalog sources are typically
listed as detections at one or more wavelength bands. NED objects are intended to represent unique astrophysical objects. 
For each object, NED provides cross-identifications to any catalog sources that NED has cross-matched to them,
their positions, redshifts, photometric data, diameter measurements, classifications, 
morphologies, and other descriptors, where available. 

\section{Methodology}

\subsection{Cross-matching VLCs with NED}

We take the following steps to cross-match a catalog in NED.
First, we load catalog source data (position, name, photometry, etc.) into the database before matching. 
At this point the names and positions of VLC sources with positions may be made immediately available for perusal in NED.
Next, we perform a positional search for match candidates in the database with the NED Cone Search (CSearch).  
Then we run the NED Match Expert (MatchEx) on a representative sample of match candidates to tune match rules and thresholds.
After statistical optimization, we run MatchEx on the entire VLC. Any matches, new objects, or associations are loaded
into the database by the Object Loader.

We use the PostgreSQL stored procedure CSearch to select all NED objects
within a fixed search radius $R_s$ of each catalog source as match candidates. The number of background sources 
is counted within a fixed background radius $R_b$, for use in computing the Poisson match probability.
We also search for neighboring catalog sources within $R_s$ of each catalog source, in order to identify 
candidate match conflicts.

We use the Python program MatchEx to select the best match candidate (if any) to each catalog source. MatchEx operates
on source and object parameters from CSearch output. MatchEx currently uses source and object positions, position uncertainties,
separation $s$, types (e.g. UvS, QSO), names, background object density n, and telescope beam size. The separation
uncertainty $\sigma$ is taken to be the sum in quadrature of the catalog and NED position uncertainties. MatchEx uses
conditional logic to determine which NED objects in the search region centered on each catalog source are acceptable matches.
The match criteria include thresholds on separation $s$, normalized separation $r = s/\sigma$, Poisson probability $P$,
and type and name preferences or exclusions. Additional criteria for future VLC matching may utilize photometry, redshifts,
and detailed morphological or spectral classifications.

For any given catalog source there are three possible MatchEx outcomes. If one object meets all match criteria then it makes the
cross-ID. If no single object meets all match criteria then it creates a new NED object and any associations. If multiple (N) sources
match a single object, then it creates N new objects and (N+1)N/2 associations.

\subsection{Position-based Match Probability}

MatchEx uses the Poisson statistic to estimate the false-positive rate for pure position 
matching. However, this value should be regarded as an upper limit since the MatchEx selection algorithm
uses additional source and object parameters to eliminate background and improve match accuracy. 
The number $N$, and mean local surface density $n=N/(\pi R_b^2)$ of NED objects is measured within the background 
radius $R_b$, and used to estimate the background contamination rate. The Poisson 
probability is computed from the Poisson distribution $P_s(x=k)=<N_s>^k exp(-<N_s>)/k!$, where $x$ is the number of 
sources found within separation $s$, and $<N_s>= n \pi s^2 = N(s/R_b)^2$ is the expected number of background objects
within $s$. For each source-object match candidate, we compute the likelihood that $k=0$ background sources are found 
closer to the source than $s$, $P=P_s(x=0)=exp(-<N_s>)=exp(-N(s/R_b)^2)$. Summing up the Poisson probabilities 
for all matches gives the false-positive match rate $f_p = (1-\sum P)/N_G/100$. Note that the $f_p$ value has to be determined 
experimentally from the MatchEx matching results. We can tune the false-positive match rate by raising or lowering the Poision 
probability match threshold $P_t$.

For the most efficient search, the ratio $R_b/R_s=$(background radius)/(search radius) needs to be adjusted to the Poisson probability
threshold. If this ratio is too small, objects found in an outer annulus of the search region will have Poisson probabilities below 
the threshold, even when there are zero additional objects inside the background radius. This ratio is optimal for $N=1$ background 
sources when $P=exp(-1\times(R_s/R_b)^2)=P_t$. For example, $R_b/R_s=3.09$ is the most efficient value to use for 
a Poisson threshold of 0.90.

\subsection{Match Selection Logic}

The match selection logic used by MatchEx is illustrated in Figure 1 and summarized as follows. A match is made between a catalog 
source and a NED object when there is a single object in the search radius with $s<=s_\mathrm{cut}$, $r=s/\sigma<=r_\mathrm{cut}$, 
$P>=P_\mathrm{cut}$, of allowed object type, and that does not overlap another NED object. Overlap occurs when two or more NED objects have 
overlapping position error circles (95\% enclosed probability), such that there is a significant chance that they represent 
the same astrophysical object. If there are multiple objects matching the above criteria, and only one of them
has a preferred object type or preferred cross-ID prefix, then that object is a match. Otherwise, if there 
are multiple objects matching the above criteria, a match is made to the closest object if the ratio of separation to 
the second closest object is $s_1/s_2<0.33$. However, if there are two or more objects of the same type within 
$s<=1.0\arcsec$, these are assumed to be duplicate NED objects, and the source will be matched to the object with the most 
cross-ID's (a measure of popularity).

\begin{figure}[!ht]
\plotone{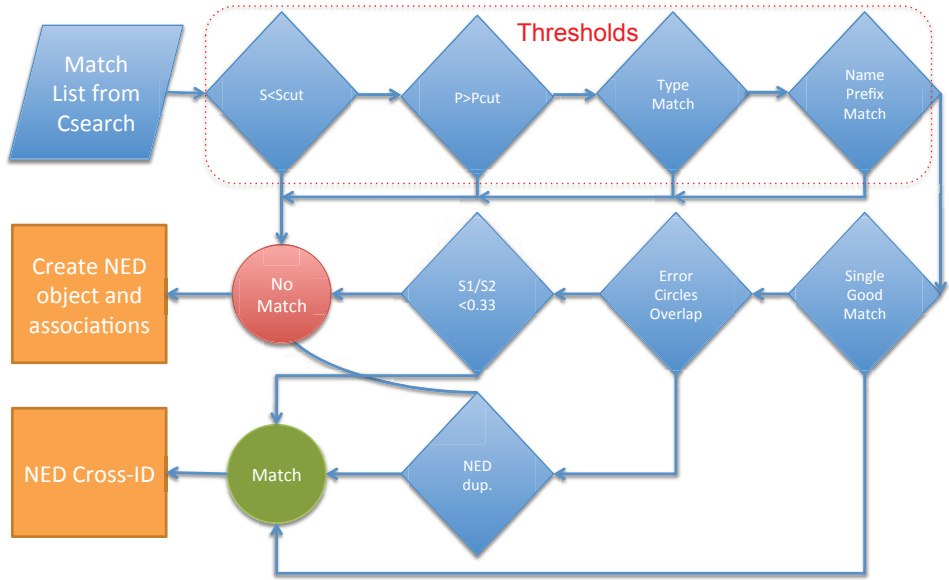}
\caption{MatchEx match selection logic.}
\end{figure}

\subsection{Preferred and Excluded Object Types and Names}

A good way to reduce the background object confusion is to exclude illogical,
problematic, or unlikely object types or names, or to preferentially match preferred (likely) object
types or names.  For GALEX ASC, objects with AbLS, GClstr, and GGroup types are excluded because 
they should not match to a (single) optical source.  A generic GammaS, X-rays, SmmS, or RadioS that has 
not previously been cross-ID'd with an optical source will typically have a large beam and large positional uncertainty 
not suitable for matching.  Matches to variable sources such as Novae and SNe were not allowed since source observation 
times are not included in the catalog. GASC source matches to NED objects with object types or cross-ID types of UvES (UV-excess) 
or with object name or cross-ID name beginning with the string ``GALEX'' were preferred. This means that if there was only
one UvES or GALEX match candidate within the search radius and it fell within the match thresholds for
all other parameters, it was selected.

\subsection{Associations}

Two types of associations are created to indicate a non-match relationship between catalog sources and NED objects:
ErrorOverlap and InBeam. An ErrorOverlap association is created where a catalog source has a position error circle that 
overlaps that of a NED object and vice versa. This is an example of a symmetric association ($\leftrightarrow$).  An InBeam association
indicates a NED object that falls inside a catalog source beam ($s<4.8\arcsec$ for GALEX). This is an asymmetric association ($\rightarrow$),  
useful for indicating when a catalog source may combine photons from  multiple NED objects, or if a candidate match was rejected for other 
reasons, even though it was the only NED object that fell inside the source beam.

\section{Input Data}

The GALEX All-Sky Survey Source Catalog \citep[GASC;][]{s12} 
contains 39,570,031 NUV-selected sources, corresponding to $>3\sigma$ detections in the NUV band . The GASC survey 
covers 26,300 square degrees (8.0 sr, or 63.8\% of the sky), consisting of all GALEX exposures with exposure times
$<800$ sec (typically 100 sec), imaged to a mean depth of NUV$=20.5$ mag (AB). Gaps in sky coverage include the 
Galactic plane, Magellanic Clouds, and regions containing bright stars.  The GALEX imaging FWHM is $5.3\arcsec$ in the
NUV band, giving a Gaussian beam diameter of $9\farcs6$ at 10\% peak flux. GASC Photometry consists of measurements in the 
NUV ($\lambda=2316$\AA, $\Delta\lambda=1000$\AA) and FUV ($\lambda=1539$\AA, $\Delta\lambda=400$\AA) bands. NED has selected 
photometry from two different methods for each of the FUV and NUV bands to include in its photometric database. The first method 
gives the Kron flux in an elliptical aperture, which is appropriate for extended sources. The second method gives the flux in a 
$7.5\arcsec$ diameter aperture.
GASC was matched to candidates selected from the NED 23.7 production database in 2013 October. That version of NED 
contained roughly 180 million unique objects derived from  $90,211$ references, with $222$ million multiwavelength cross-IDs.

\begin{figure}
\plotone{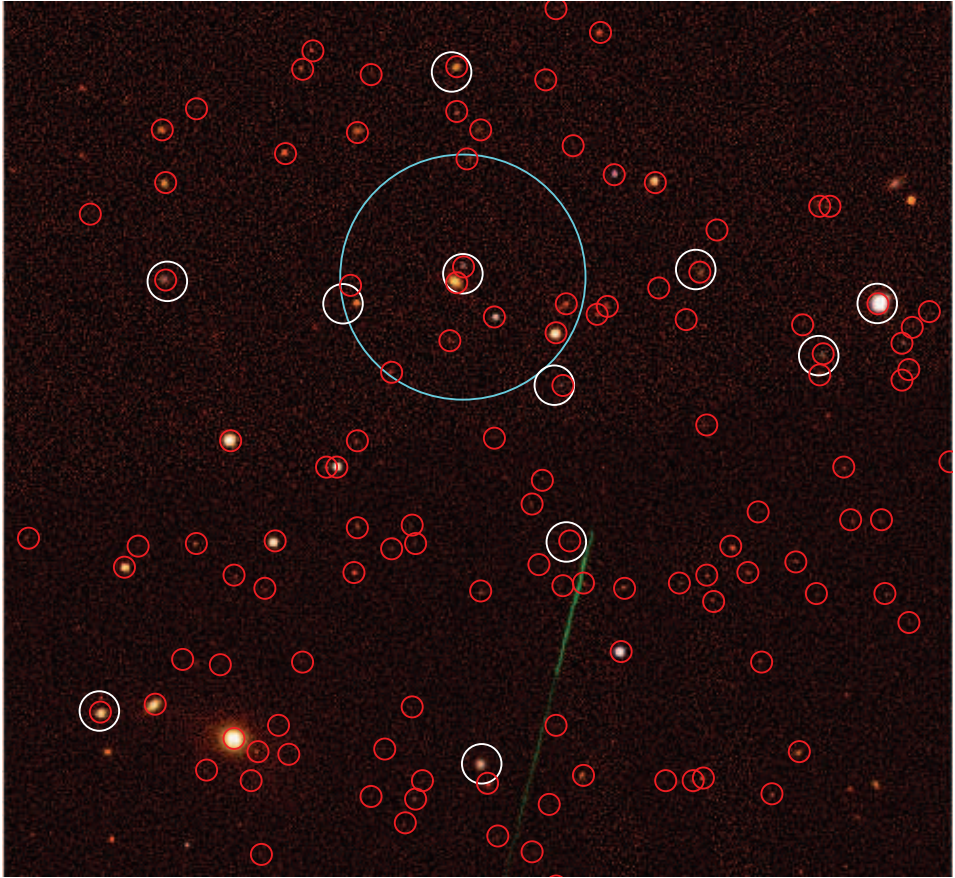}
\caption{Overlay of $7\farcs5$ radius search regions for GALEX ASC NUV sources (white circles) on $6\arcmin \times 6\arcmin$ SDSS DR6 gri image, centered at (RA,Dec)=(200,+30).
One of the regions used to estimate the local background object density surrounding one of the sources is shown in cyan. The locations of NED objects are indicated by red circles. }
\end{figure}

\section{Tuning MatchEx}

To tune the MatchEx algorithm, match thresholds, and performance, test runs were conducted using $10^5$ GASC sources in 
circular regions centered at (RA,Dec)$=$(200,+30) in the Sloan Digital Sky Survey (SDDS) North Galactic Pole (NGP) survey region; 
and outside the SDSS survey region, centered at (RA,Dec) $=$ (60,+30).  The search radius for each source (Figure 2) was selected to be $R_s=7.5\arcsec$ 
after considering the observed separation distribution of  GASC-NED matches. 
A background radius of $R_b=6.2 R_s =46\farcs5$ was used for measuring the local background density for use in calculating the Poisson match likelihood.

The goal in tuning the match thresholds is to minimize both the false positive $f_p$ and false negative $f_n$ match rates. These two rates
tend to offset one another, with more strict thresholds reducing the false positive rate at the expense of increased false negative
rate. Where this trade-off is optimized depends on how much weight is given to accuracy versus completeness. Because we think it is much 
worse to make an incorrect match than to miss a potentially good match, we use the combined error rate $f_e=f_n+10\times f_p$ as our
performance metric for  MatchEx. 

The Poisson statistic is the primary statistic that we use to determine match likelihood, and it is directly related to the false positive
match rate. We ran the GASC 100K test several times with Poisson statistic threshold in the range $P_t=0.82-0.98$ to find the dependence of 
$f_p$, $f_n$, and $f_e$ on $P_t$ (Figure 3).  We find that $f_e$ is minimized for a Poisson threshold of $P_t=0.90$. The minimum in $f_e$ is rather
broad, so the precise value of $P_t$ does not make much difference in the range $P_t=0.88-0.92$. We have not yet made a detailed
study of the impact of other thresholds and selection criteria (search radius, object type exclusions, and object name preferences) on 
the match error rate.

\begin{figure}
\plotone{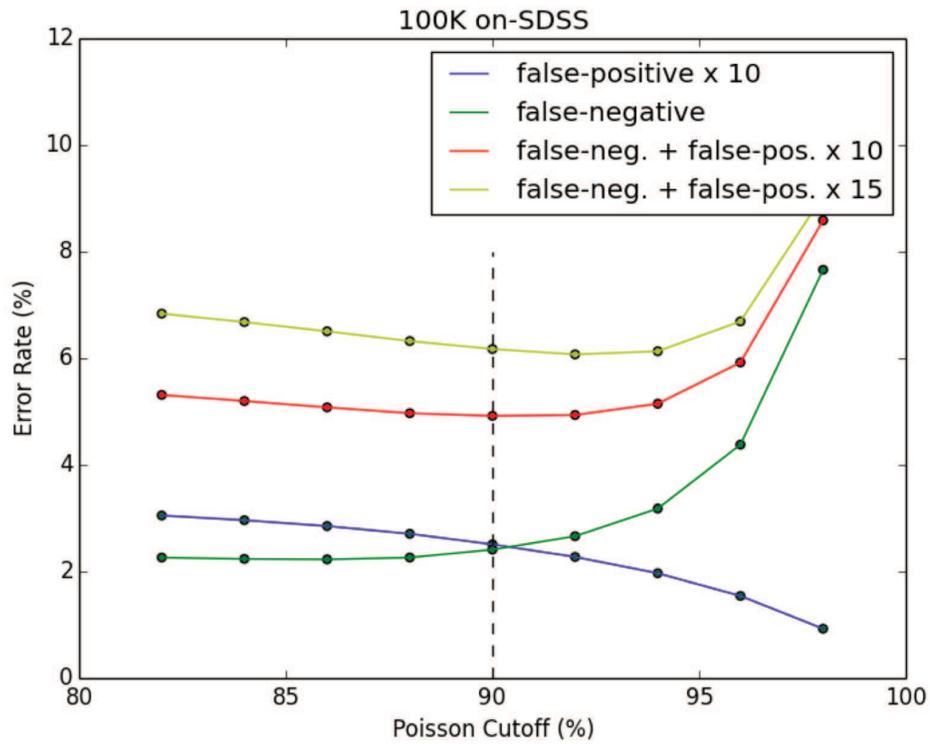}
\caption{Optimization of match error metric vs. Poisson statistic threshold $P_t$. The false positive rate drops, while the false negative rate rises with $P_t$.
The combined match error metric $f_e=f_n+10 \times f_p$ has a minimum at $P_t=0.90$. }
\end{figure}

\section{Results}

\subsection{Overall Statistics}

From the 39,570,031 GASC UV sources and 23,301,552 NED object match candidates, there are 10,595,382 (26.8\%) matches 
to NED objects and  28,974,649 (73.2\%) no matches, of which 26,984,670 (68.2\%) are no-matches in NED blank fields. 
The remaining 1,992,979 (5.0\%) of no-matches occur in non-blank fields, including fields with one match candidate 
(2.6\%), two match candidates (1.2\%), 3 match candidates (0.7\%), and 4 or more match candidates (0.6\%).

We present distributions for the position-based matching parameters in Figure 4 for MatchEx selected 
matches, compared to the unfiltered CSearch selected match candidates.  The number of match candidates is expected to increase linearly 
with separation at large separation ($s>4\arcsec$), for uniform mean background density: $N = <n> \pi s^2$,  $dN/ds = 2\pi <n> s$. 
By integrating under the linear background fit, we estimate the number of background objects within $s_t=7.5 \arcsec$ to be 
2,980,000 (21.5\% of match candidates). This is similar to the number of rejected match candidates (3,275,249 or 23.6\%) inside 
this radius. The actual match separation distribution is close to the one obtained by subtracting the linear background fit from the match 
candidate separation distribution, indicating that MatchEx does a good job of eliminating background NED sources as matches. 
The small difference between the two at $s=3.5-7\arcsec$ gives an estimate of$f_n=2.4\%$ for the false negative match rate.

The distribution of $dN/dr$ vs. $r$ is compared to the derivative of a Gaussian function (Fig. 4). The peak of the $dN/dr$ vs. 
$r$ distribution lands near $0.9\sigma$, showing that the scale of the combined GASC and NED position error bars may be
overestimated by 10\%. However, there is an excess of matches at $r>1.5\sigma$ compared to a Gaussian with standard 
deviation $\sigma_r=0.9\sigma$. This shows that the error distribution in separation is not perfectly matched to a 
Gaussian distribution. We used a match threshold of $r=s/\sigma=3.5$, aiming to limit the match incompleteness to 0.2\%. 
However, the non-Gaussian distribution of separation errors led to greater incompleteness. 

The Poisson probability density distribution for CSearch match candidates peaks sharply at a value of $P=1$. There is a 
tail of match candidates with probabilities $0<P<0.9$, corresponding to background NED objects. The chosen Poisson match 
threshold of $P_t=0.9$ roughly matches the location in the distribution where the density of true matches begins to exceed 
the number of false-positives. 

\begin{figure}
\plotone{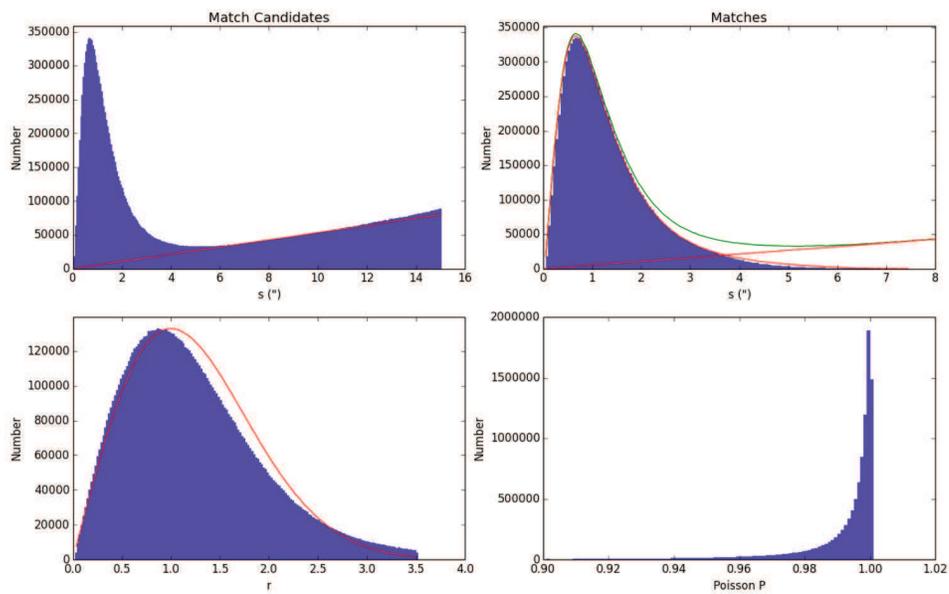}
\caption{Distribution of separation $s$, normalized separation $r=s/\sigma$, and Poisson match statistic $P$ for GASC-NED matches
              The linear background estimate is indicated by the red line in the upper-left panel.  The upper-right panel compares the distribution 
              of match separations to candidate matches (green line) and background-subtracted candidate matches (red line). 
              The normalized separation histogram is compared to the derivative of the Gaussian error distribution with 
              $\sigma_r=0.9,1.0$ (red curves in the lower left panel). There is significant deviation from a Gaussian  distribution of positional errors.}
\end{figure}

\subsection{Object Type Statistics}

Table 1 gives the breakdown of GASC-NED candidate matches and matches by NED preferred object type (\url{http://ned.ipac.caltech.edu/help/faq5.html#5k}). 
The top-9 most abundant types account for 99.94\% of the total matches. Galaxies (G) are most abundant, making up roughly 61\% of GASC-NED matches. 
Stellar objects (*) are second, making up roughly 36\% of matches.  The remaining 3\% of matches are to NED object types VisS, QSO, !*, UvES, IrS, WD*, and
GPair in order of abundance. To study the object type abundance patterns for matches, we take the ratio of their 
abundances divided by the overall abundances in NED (last column of Table 1). Matches to objects of type QSO, !*, UvES, and WD are  significantly enhanced, 
by factors of 9-14 relative to their NED abundances, which may reflect the tendency for these types of objects to be relatively UV-bright. On the other hand, 
matches of type IrS are under abundant, with a ratio of 0.32 with respect to NED.
 
\section{Future Improvements}

Match accuracy may be improved by considering the full array of object data compiled by NED, including redshifts, photometry, diameters,
and detailed classifications. Spectroscopic redshifts are only available for a small fraction of NED objects, but can provide a strong constraint
for matching. Photometric constraints should be relatively weak since we do not want to bias matches against objects with unusual SEDs. We 
also plan to make use of object size, overlap, and type for extended sources.  Finally, we are generalizing our cross-matching algorithm
to handle smaller, heterogeneous catalogs federated from the NED literature pipeline.

\begin{table}
\caption{Match Counts by Object Type}
\smallskip
\begin{center}
{\small
\begin{tabular}{lrrrrrrrr}
\tableline
\noalign{\smallskip}
Type & NED & Cand. & Match & Matched & NED & Cand. & Match & Abund.\\
           Code &     &       &       &   \%    & \%  &   \%  & \%    & Ratio\\

\noalign{\smallskip}
\tableline
\noalign{\smallskip}
G          &102,386,926 & 13,816,821 & 6,447,394 & 46.6 & 57.2   & 59.3  & 60.8  & 1.06\\   
$\ast$     & 68,268,375 &  8,391,557 & 3,801,913 & 45.3 & 38.1   & 36.0  & 35.9  & 0.94\\   
VisS       &  1,821,299 &    321,549 &    99,406 & 30.9 &  1.02  &  1.38 &  0.94 & 0.92\\   
IrS        &  1,661,077 &    239,803 &    31,370 & 13.1 &  0.93  &  1.03 &  0.30 & 0.32\\
RadioS     &  2,025,701 &    152,692 &         0 &  0.0 &  1.13  &  0.66 &  0    & 0\\
QSO        &    163,260 &     98,648 &    86,681 & 87.9 &  0.091 &  0.42 &  0.82 & 9.01\\
!$\ast$    &     89,416 &     69,301 &    58,859 & 84.9 &  0.050 &  0.30 &  0.56 &11.2 \\
UvES       &    129,330 &     62,752 &    56,238 & 89.6 &  0.072 &  0.27 &  0.53 &13.6 \\
XrayS      &    407,843 &     52,820 &         0 &  0.0 &  0.23  &  0.23 &  0    & 0\\
GGroup     &     92,910 &     34,739 &         0 &  0.0 &  0.052 &  0.15 &  0    & 0\\
GPair      &     26,669 &      9,759 &     2,312 & 23.7 &  0.015 &  0.042&  0.022& 1.47\\
WD$\ast$   &      9,461 &      7,951 &     7,574 & 95.3 &  0.0053&  0.034&  0.071&13.4 \\
\noalign{\smallskip}
\tableline
\end{tabular}
}
\end{center}
\end{table}

\acknowledgements The NASA/IPAC Extragalactic Database (NED) is operated by the Jet Propulsion Laboratory, 
California Institute of Technology, under contract with the National Aeronautics and Space Administration. We give special thanks to Mark Seibert 
(Carnegie Observatories) for providing a copy of the GALEX catalogs and for answering our questions about them.

\end{document}